%% file: main.tex
\documentclass[11pt]{article}

\usepackage[T1]{fontenc}
\usepackage{lmodern}
\usepackage{geometry}
\usepackage{graphicx}
\usepackage{subcaption} 
\usepackage{amsmath,amssymb}
\usepackage{booktabs}
\usepackage{longtable}
\usepackage{array}
\usepackage{float}

\usepackage[numbers]{natbib}
\usepackage[colorlinks=true,allcolors=blue]{hyperref}

\title{Program-Level Curriculum Analysis of U.S. Quantum Master’s Degrees;
Implications for Workforce Preparation}
\author{Bradley Holt}
\author{
Tunde Kushimo$^{1}$,
Bradley Holt$^{2}$,
Muhammad Talal$^{3}$\\[0.5em]
\small
$^{1}$Department of Mathematics, Statistics, and Physics, Wichita State University, Wichita, KS 67260, USA\\
$^{2}$IBM Quantum, Madison Ave, New York, NY 10010, USA\\
$^{3}$Department of Physics, Quaid-i-Azam University, Islamabad 45320, Pakistan
}

\begin{document}
\maketitle

\begin{abstract}
Quantum technologies are increasingly recognized as a strategic priority for economic competitiveness, national security, and technological innovation in the United States. As quantum systems transition from research prototypes to deployable technologies, attention has shifted toward the preparedness of the quantum workforce, particularly the alignment between higher education and industry skill needs. While prior research has examined individual aspects of quantum education or workforce demand, few studies integrate systematic curriculum analysis with documented industry expectations. This study addresses that gap by analyzing primary U.S. master’s programs in quantum science and technology, focusing on curriculum structure and skill development. Using a structured coding framework, course offerings were mapped across six quantum-relevant skill categories and aggregated to produce program-level skill profiles. These profiles were then compared with industry-identified competencies reported in recent workforce studies. The findings reveal strong emphasis on quantum theory across programs, alongside substantial variability in technical skills, applied learning opportunities, and professional development components. The results highlight areas of alignment as well as persistent gaps related to workforce readiness, cross-disciplinary integration, and emerging technological demands. This study provides a scalable framework for evaluating quantum education programs and offers evidence-based insights for curriculum design, workforce policy, and the continued development of the U.S. quantum ecosystem
\end{abstract}

\section{Motivation}

Quantum technologies, including quantum computing, communication, sensing, and cryptography, are widely considered the foundation of the next technological revolution. Governments, industries, and research institutions around the world have made significant investments to accelerate the development and deployment of quantum systems, motivated by their potential to solve classically intractable problems, enable secure communications, and transform precision measurement. In the United States, these investments are accompanied by a growing recognition that the availability of a highly skilled workforce is a critical determinant of long-term success in quantum innovation.

As the quantum ecosystem matures, concerns have emerged regarding the alignment between academic training and industry workforce needs. Traditional quantum education has been rooted primarily in physics-based instruction, emphasizing mathematical formalism and theoretical foundations. While this grounding remains essential, the modern quantum industry increasingly demands interdisciplinary competencies spanning computer science, engineering, applied science, mathematics, systems integration, and other related domains, alongside professional skills including communication and project management. Industry reports consistently note challenges in recruiting talent that combines quantum knowledge with practical, workforce-ready skill sets, particularly for non-PhD roles.

In response, universities have expanded quantum-focused educational offerings, including certificate, minor, and master’s level programs, deliberately positioning these pathways as primary entry points into the quantum workforce. However, existing research on quantum education in the United States has largely focused on descriptive research of course offerings, pre-college initiatives, or broad workforce needs, often treating education and industry demand as separate domains. As a result, there remains limited empirical understanding of how well current U.S. quantum programs align with domestic industry requirements, where gaps persist, and how programs vary in their approaches to interdisciplinarity and applied learning.

This study addresses these gaps by conducting a systematic analysis of primary U.S. quantum master’s programs. Using a structured curriculum coding methodology, the study examines how programs distribute instructional emphasis across key quantum skill categories, the extent to which applied learning and professional development are embedded, and how these educational outcomes align with industry-identified competencies. By integrating curriculum data with established workforce literature, the study aims to provide actionable insights for educators, policymakers, and industry stakeholders seeking to strengthen the U.S. quantum talent pipeline.

Since this study was first conducted in 2025, the number of quantum programs in the United States has continued to grow. As of May 2026, 24 quantum-focused programs are offered across 19 U.S. institutions, underscoring the rapid expansion of the quantum education ecosystem and the urgency of examining curriculum–workforce alignment.

\subsection{Literature Review}

The development of a quantum-ready workforce has become a central focus of research at the intersection of quantum science, education, and workforce development. Early scholarship emphasized the strategic importance of human capital in sustaining quantum innovation, highlighting the need for coordinated education pathways that extend beyond traditional physics training (Aiello et al., 2021; Fox et al., 2020). These studies established the foundational argument that workforce preparedness, rather than technological capability alone, would shape the pace of progress in quantum technologies.

Subsequent research has examined the structure and content of quantum education programs, particularly in the United States. Surveys of undergraduate and graduate coursework reveal that quantum information science education is heavily concentrated in physics departments, with curricula emphasizing quantum mechanics, quantum information theory, and mathematical foundations (Cervantes et al., 2021; Meyer et al., 2024). Although these programs increasingly incorporate topics such as quantum computing and algorithms, the degree of interdisciplinarity varies widely across institutions. This variability reflects both institutional autonomy and the absence of a unified national framework guiding the design of quantum curricula.

In parallel to education-focused studies, a growing body of literature has documented industry perspectives on quantum workforce needs. Industry surveys and needs assessments consistently identify the demand for skills beyond theoretical knowledge, including quantum programming, software development, hardware familiarity, experimental techniques, and systems-level thinking (Hughes et al., 2022; Dandridge, 2023; Piña et al., 2025, Shams El-Adawy et al., 2025). Importantly, these studies emphasize the growing importance of roles that do not require doctoral-level training, underscoring the relevance of master’s programs as workforce pipelines. However, industry reports often lack direct mapping to educational curricula, limiting their utility for program-level evaluation.

More recent work has attempted to bridge education and workforce perspectives by proposing competence-based frameworks for quantum education. European initiatives, in particular, have advanced structured competence models that define skill profiles across quantum subfields and qualification levels (Greinert et al., 2023; Greinert et al., 2024). These frameworks provide a common language for aligning education, training, and industry needs; however, their applicability to the U.S. context, characterized by decentralized governance and institutional diversity, remains an open question.

Applied learning and experiential education have also emerged as recurring themes in the literature. Studies highlight internships, capstone projects, and research placements as critical mechanisms for developing workforce-relevant competencies and professional identity (Hasanovic et al., 2022; Plunkett et al., 2020). Yet empirical evidence suggests that access to such opportunities is uneven across programs, potentially exacerbating disparities in student preparedness and career outcomes.

Finally, research examining student perspectives reveals persistent challenges related to career awareness and workforce transition. Many students report limited understanding of available quantum career pathways, particularly in industry settings, and express uncertainty about how academic training translates to professional roles (Rosenberg et al., 2024; Oliver et al., 2025). These findings suggest that curriculum content alone is insufficient to ensure workforce readiness without explicit attention to professional development and career signaling.

Collectively, the literature underscores the need for integrated, data-driven analyses that connect quantum education structures with industry workforce demands. While prior studies provide valuable insights into individual components of the quantum ecosystem, few systematically examine curriculum-level skill distributions across institutions or assess alignment with documented industry needs within a unified analytical framework. This study builds on existing scholarship by addressing these gaps through a structured analysis of U.S. quantum master’s programs, directly informing questions of curriculum design, workforce policy, and the future of quantum education in the United States.

\subsection{Research Questions}

Building on the gaps identified in the literature and the need for systematic alignment between quantum education and workforce demands, this study is guided by the following research questions:

\textbf{RQ1.} How are current U.S. quantum master’s programs structured in terms of curriculum content and skill development?
This question examines how instructional emphasis is distributed across key quantum skill domains, including theoretical foundations, applied technologies, and professional competencies. It focuses on identifying patterns and variation in curriculum design across institutions, providing a comparative view of how programs conceptualize preparation for the quantum workforce.

\textbf{RQ2.} To what extent do the competencies developed in U.S. quantum master’s programs align with the skill needs identified by domestic quantum industry stakeholders?
This question evaluates alignment between program-level skill profiles and industry-reported workforce needs. By comparing aggregated curricular skill distributions with documented industry demand, the study assesses areas of convergence as well as divergence between education and employment expectations.

\textbf{RQ3.} What gaps exist between educational outcomes and industry requirements, particularly with respect to cross-disciplinary skills, workforce readiness, and emerging quantum technologies?
This question synthesizes findings from the first two questions to identify persistent gaps in curriculum coverage, applied learning, and professional preparation.

Together, these research questions provide a structured framework for analyzing the current state of U.S. quantum master’s education, evaluating its alignment with industry needs, and identifying opportunities for curriculum innovation and policy intervention.

\clearpage
\section{Methodology}

Many countries have recognized the importance of master’s-level education by explicitly emphasizing such programs within their national quantum strategies, including those of the United States. These strategies increasingly position quantum technology master’s programs as a scalable and timely pathway for preparing a quantum-ready workforce.

In this study, a quantum technology master’s program is defined as a one- or two-year degree-granting program focused on quantum science, quantum engineering, or closely related quantum technologies. This study employs a systematic, document-based curriculum analysis to examine the extent to which U.S.-based quantum-focused master’s programs align with contemporary quantum industry workforce skill requirements. The methodology integrates qualitative coding, quantitative normalization, discretization-based aggregation, and robustness analysis to generate comparative program-level and system-level assessments of skill emphasis, and applied workforce preparation.

The scope of this study focuses on primary, degree-granting master’s programs in quantum science, quantum engineering, or closely related quantum technologies offered by U.S. institutions. Following the definition of Goorney et al. (2025), primary master’s programs are those in which the entire curriculum is structured around at least one of quantum computing, quantum sensing, quantum communication, quantum simulation, or quantum engineering, as reflected in the degree title. Program selection followed criteria consistent with prior mappings of the U.S. quantum education landscape, prioritizing programs that explicitly identify quantum science or technology as the central organizing theme rather than as a specialization embedded within a traditional physics or engineering degree.

The identification of primary master’s programs involved compiling existing, non-exhaustive listings from prior mappings of the U.S. quantum education landscape, which were subsequently updated through systematic internet searches conducted on a state-by-state basis. This process resulted in the identification of fifteen primary quantum master’s programs across the United States, updating the number of programs identified by Goorney et al. (2025); all fifteen programs were included in the final selection. The selected programs span a range of institutional types, geographic regions, and disciplinary orientations, including physics-led, engineering-led, and explicitly interdisciplinary programs. Collectively, these programs are intended to capture the dominant curricular models currently shaping graduate-level quantum workforce preparation in the United States. Table~\ref{tab:us_quantum_masters} provides an overview of the quantum master’s programs included in the study.

\begin{table}[h]
\centering
\caption{U.S. Primary Quantum Master’s Programs}
\label{tab:us_quantum_masters}
\begin{tabular}{lll}
\toprule
UNIVERSITY & Abbrev. used & QUANTUM PROGRAM (MS) \\
\midrule
University of Delaware & Delaware & Quantum Science \& Engineering \\
University of California, Los Angeles & UCLA & Quantum Science and Technology \\
University of Maryland & Maryland & Quantum Computing \\
University of Wisconsin--Madison & Wisconsin & Quantum Computing \\
Southern Methodist University & SMU & Quantum Engineering \\
University of Southern California & USC & Quantum Information Science \\
Stony Brook University & Stony Brook & Quantum Information Science and Technology \\
Colorado School of Mines & Colorado & Quantum Engineering \\
University at Buffalo (SUNY) & SUNY & Quantum Science and Nanotechnology \\
Indiana University Bloomington & Indiana & Quantum Information Science \\
San José State University & San Jose & Quantum Technology \\
Stevens Institute of Technology & SIT & Quantum Engineering \\
University of Rhode Island & Rhode Island & Quantum Computing \\
Columbia University & Columbia & Quantum Science and Technology \\
Rutgers University & Rutgers & Quantum Science\\
\bottomrule
\end{tabular}
\end{table}

\subsection*{Data Collection and Validation}

Program-level information was collected through a systematic review of publicly available institutional materials between October and November 2025. Sources included official program websites, degree requirement pages, university course catalogs, program handbooks and curriculum outlines, detailed descriptions of required and elective courses, and publicly advertised experiential learning components such as internships, research laboratories, and capstone projects. For each program, a comprehensive list of all required and elective courses was compiled. To ensure accuracy, course information was cross-checked across multiple sources when available, and discrepancies were resolved through consensus among the research team. Restricting data collection to publicly available materials ensured transparency and replicability of the analysis.

Only information explicitly documented in official materials was coded. Anecdotal program elements not described in public sources were excluded from the analysis.

\subsection{Coding Framework}

The compiled curricular data were systematically coded using a predefined skill taxonomy to classify courses according to their emphasis on core quantum topics, interdisciplinary competencies, and applied workforce skills. The coding scheme was designed to capture the presence and relative emphasis of skill domains within program curricula rather than instructional depth, rigor, or specific learning outcomes.
\\
\\
\textit{The taxonomy consists of the following skill categories:}
\\
\\

\textbf{Quantum Theory/ Information Science: (QTheory)}
This category captures the foundational principles and mathematical framework underlying quantum mechanics and quantum information processing. Representative topics include quantum states, superposition, entanglement, measurement theory, density matrices, operator formalism, many-body quantum systems, quantum statistical mechanics, complexity theory, quantum error correction, and spin systems.

\textbf{Quantum Hardware/ Engineering Devices: (QHard)}
This category encompasses the design, fabrication, characterization, and optimization of physical systems and materials for quantum technologies. Representative topics include quantum hardware platforms (solid-state, photonic, atomic, and superconducting systems), device operation, fabrication techniques (e.g., lithography and thin-film deposition), cleanroom processes, laser spectroscopy, cryogenic systems, semiconductor and superconducting devices, qubit implementation platforms, and experimental techniques for controlling and measuring quantum systems. Particular emphasis was placed on courses incorporating hands-on or laboratory-based experiences.

\textbf{Quantum Algorithms/ Software/ Quantum Machine Learning: (QSoft)}
This category includes the design, implementation, and analysis of algorithms for quantum computation, as well as practical software frameworks for quantum programming. Representative topics include quantum algorithms (e.g., Grover’s and Shor’s algorithms), variational and hybrid classical–quantum algorithms, quantum simulation, optimization, quantum machine learning, VQE, QAOA, quantum gates and circuits, quantum programming languages, linear algebra for quantum computing, quantum annealing, and quantum teleportation.

\textbf{Quantum Networking/ Communication/ Cryptography: (Qcom)}
This category addresses secure information transfer and networking using quantum principles. Representative topics include quantum key distribution protocols, entanglement-based communication, quantum teleportation, dense coding, quantum channels, quantum repeaters, entanglement distribution, error correction, cryptography, and classical–quantum hybrid communication frameworks, including foundational concepts from information theory.

\textbf{Quantum Sensing/ Metrology: (QSense)}
This category covers techniques for high-precision measurement and detection using quantum systems, leveraging quantum coherence, entanglement, and superposition to surpass classical limits. Representative topics include quantum sensors, measurement protocols, quantum metrology, interferometry, atomic clocks, magnetometry, optical detection, photon counting, precision measurement, signal processing, noise modeling, and decoherence analysis.

In this study, \emph{communication/project management} and \emph{career awareness} were treated as distinct non-technical skill categories and operationalized strictly based on the presence of \emph{stand-alone courses} whose stated instructional objectives explicitly focused on these competencies. Courses coded under \emph{communication/project management} were limited to dedicated offerings designed to train students in scientific communication and/or project management practices, rather than technical courses that incidentally incorporated group projects or presentations. These stand-alone courses emphasized transferable professional competencies required for collaborative and interdisciplinary technical work, including project planning and execution, resource allocation, risk management, team-based coordination, technical writing and oral communication, presentation of results, and organizational or leadership practices relevant to complex scientific and engineering contexts. No inference of communication or project management skill development was made from technically focused courses unless such competencies constituted the primary and explicit instructional focus of the course.

Figure 1 presents a representative sample of courses, illustrating how course content was categorized into primary skill areas for our analysis. Some courses were associated with one or two primary skill categories, while others included multiple secondary skill categories.

\begin{figure}[htbp]
\centering
\small
\renewcommand{\arraystretch}{1.3}
\begin{tabular}{p{5cm} p{9cm}}
\hline
\textbf{Sample Course (Core Skill)} \\
\hline
Intro Quantum Computation \& Information (QTheory) &
Foundations of quantum computation and information, including qubits, measurement, entanglement, decoherence, quantum gates, algorithms, cryptography, quantum key distribution (QKD), and quantum error correction. \\

Quantum Programming (QSoft) &
Practical introduction to quantum computing using Python-based quantum programming frameworks. Covers simulation of quantum algorithms, variational quantum eigensolvers (VQE), optimization, finance applications, quantum machine learning (QML), and execution on IBM quantum hardware and quantum annealers. \\

Semiconductor Device Design \& Fabrication (QHard) &
Design and fabrication of bipolar and MOS integrated circuits, including photolithography, diffusion, metallization, and device testing. \\

Theory of Cryptography (QCom) &
Design and analysis of cryptographic systems, including number theory, statistics, symmetric and asymmetric encryption, quantum encryption, RSA, hashing, digital signatures, key management, and steganography. \\

Active RF \& Microwave Devices (QSense) &
RF and microwave circuits, transistors, amplifiers, mixers, oscillators, MMICs, CAD-based modeling, fabrication, and vector network analyzer (VNA) measurements. \\

Project Management (Communication / Project Management) &
Project planning, scheduling, resource allocation, optimization, control, and risk management in technical and interdisciplinary settings. \\

Quantum Engineering Seminar (Career Awareness) &
Industry-focused seminar series covering quantum computing, sensing, materials, and communications, with emphasis on professional development and career readiness. \\
\hline
\end{tabular}
\caption{Representative coursework mapped to primary quantum workforce skill categories.}
\label{fig:course-skill-mapping}
\end{figure}

Each course was coded using a discrete ordinal scale reflecting the degree of emphasis on each skill category:
\begin{itemize}
    \item 0 — Skill not present
    \item 1 — Skill partially addressed or included as a secondary component
    \item 2 — Skill constitutes a primary or dominant focus of the course
\end{itemize}

We chose a discrete ordinal scale to balance analytical rigor with the inherent limitations of publicly available curricular descriptions. Course catalogs and program materials often vary in level of detail and terminology; an ordinal coding scheme provides robustness against such variability while enabling consistent cross-program comparisons. This approach prioritizes interpretability and reproducibility, allowing emphasis patterns to be identified without over-inferring instructional depth or learning outcomes.

\subsection{Program-Level Aggregation and Normalization}

For each program, raw skill points were aggregated by summing coded values across all required and elective courses. To enable meaningful comparison across programs with differing course counts and curricular structures, aggregated skill scores were normalized into program-weighted percentage distributions.

For a given program, the percentage contribution of each skill category was calculated as:

\begin{equation}
\text{Skill Percentage} =
\frac{\text{Total Skill Points for Category}}
{\text{Sum of All Skill Points Across Categories}} \times 100\%.
\end{equation}

This normalization ensures that each program’s skill profile reflects relative curricular emphasis rather than absolute program size.

\subsection*{Discretization of Skill Emphasis}

To facilitate aggregate comparison across the full set of programs, normalized skill percentages were transformed into a discretized ordinal representation. For each program and skill category, percentage values were mapped onto a three-level scale:

\begin{itemize}
    \item 0 -- Minimally represented skill
    \item 1 -- Moderately represented skill
    \item 2 -- Strongly emphasized skill
\end{itemize}

The baseline discretization thresholds were defined as follows:

\begin{itemize}
    \item 0: 0--4.4\%
    \item 1: 4.5--19.4\%
    \item 2: $\geq$19.5\%
\end{itemize}

This discretization abstracts fine-grained percentage differences in favor of identifying structural patterns of curricular emphasis across programs.

\subsection*{Aggregate Skill Distribution Analysis}

Discretized skill values were aggregated across all fifteen programs to compute system-level skill distributions. For each skill category, aggregate emphasis was calculated as the percentage of the maximum possible score, corresponding to all programs strongly emphasizing that skill. This analysis provides a macroscopic view of how U.S. primary master’s programs in quantum science collectively prioritize different technical competencies.

\subsection*{Robustness and Sensitivity Analysis}

To evaluate the stability of aggregate findings with respect to discretization choices and subjective interpretation inherent in ordinal coding, a robustness (sensitivity) analysis was conducted using multiple alternative threshold schemes. In addition to the baseline discretization thresholds, several alternative schemes ranging from conservative to liberal were applied to the same normalized percentage data.

Across all schemes, the underlying 0,1,2 coding structure was preserved, while the threshold cutoffs defining low, moderate, and high skill emphasis were systematically varied. Aggregate skill distributions were recomputed under each scheme and compared using overlay visualizations to assess whether observed program-level and system-level patterns were sensitive to threshold selection or structurally robust.

Across reasonable variations in discretization assumptions and minor perturbations in course-level coding, the resulting aggregate distributions were found to be qualitatively stable. This indicates that the observed patterns of skill emphasis, interdisciplinarity, and workforce alignment are not driven by individual coding decisions or specific threshold choices. The robustness of these results supports the validity of the comparative conclusions drawn from the curriculum mapping.

\subsection*{Experiential Learning and Professional Preparation}

In addition to technical skill analysis, each program was coded for the presence of formal experiential and professional development components relevant to workforce preparation. These included applied learning opportunities (industry internships, research labs, and capstone or project-based courses), communication and project management training, and career awareness or professional seminar components.

These features were coded using a binary scheme (0 = absent, 1 = present) based on explicit mention in program materials. Aggregate frequencies were calculated to assess the prevalence of workforce-oriented structures across programs.

\subsection{Data Analysis and Visualization}

Quantitative analysis and visualization were used to synthesize and compare results across programs and analytical layers. Visualization techniques included bar charts to display program-level skill distributions, heatmaps illustrating relative skill emphasis across programs, aggregate bar charts summarizing system-level skill emphasis, and overlay line plots for sensitivity analysis across discretization schemes.

\subsection*{Methodological Limitations}

This methodology provides a high-level mapping of curricular emphasis rather than an evaluation of instructional quality, learning outcomes, or graduate competence. The analysis relies exclusively on publicly available program materials, which introduces several limitations. Course descriptions vary widely in level of detail and may employ institution-specific terminology, potentially affecting the consistency of skill classification. In addition, publicly available descriptions may not reflect recent curricular updates, and the absence of access to course syllabi, assessment materials, or instructional practices limits insight into instructional depth, rigor, and student experience.

The use of discretized ordinal coding necessarily abstracts fine-grained differences between courses. However, this abstraction is intentional and appropriate given the variability and incompleteness of catalog-level data. The analysis is therefore designed to identify the presence and relative emphasis of skill themes across programs rather than to measure proficiency, pedagogical effectiveness, or mastery.

\clearpage
\section{Results}
\input{Text/0.1-Big_Figures}
This section reports the outcomes of a curriculum-based skill mapping analysis conducted across fifteen U.S. primary master’s programs in quantum science and technology. The results describe observed patterns in curricular skill emphasis derived from program documentation, without inferring educational effectiveness or workforce outcomes.

Findings are presented at multiple levels of aggregation, including (i) distributions of skill categories across individual programs, (ii) cross-program comparisons of technical skills, applied learning, and professional skill integration, and (iii) system-level aggregates of technical and non-technical skill emphasis obtained through discretization-based analysis. Where applicable, robustness and sensitivity analyses are used to assess the stability of aggregate patterns with respect to methodological choices.

This section is intentionally descriptive in nature. Interpretation of these patterns in relation to industry workforce needs, alignment with external reports, and implications for quantum education and training pathways are reserved for the Discussion section.

\subsection{Program-Level Skill Distributions}

Figures 2a–2o present normalized skill distribution bar charts for each of the fifteen programs included in the study. Each figure shows the relative emphasis placed on five technical skill categories; quantum theory and information theory (QTheory), quantum hardware and engineering (QHard), quantum algorithms, software and quantum machine learning (QSoft), quantum communication and networking (QCom), and quantum sensing and metrology (QSense) with values normalized to sum to 100\% within each program.

Across all programs, quantum theory constitutes a substantial portion of curricular emphasis, typically representing between approximately 25\% and 50\% of total coded skill weight. However, the balance between theory and applied or specialized skill domains varies considerably. Some programs exhibit a strongly theory-centered profile, while others allocate comparable or greater emphasis to hardware, software, or application-driven skills.

Several programs demonstrate relatively balanced distributions across multiple technical domains, while other programs show pronounced specialization, with one or two skill categories accounting for a majority of curricular emphasis. QCom and QSense skills appear at low levels across most programs, with only a small subset allocating a notable fraction of coursework to this domain.

\subsection{Cross-Program Comparison of Skill Distribution}

\begin{figure}[H]
    \centering
    \begin{minipage}{0.45\textwidth}
        
        \includegraphics[width=\linewidth]{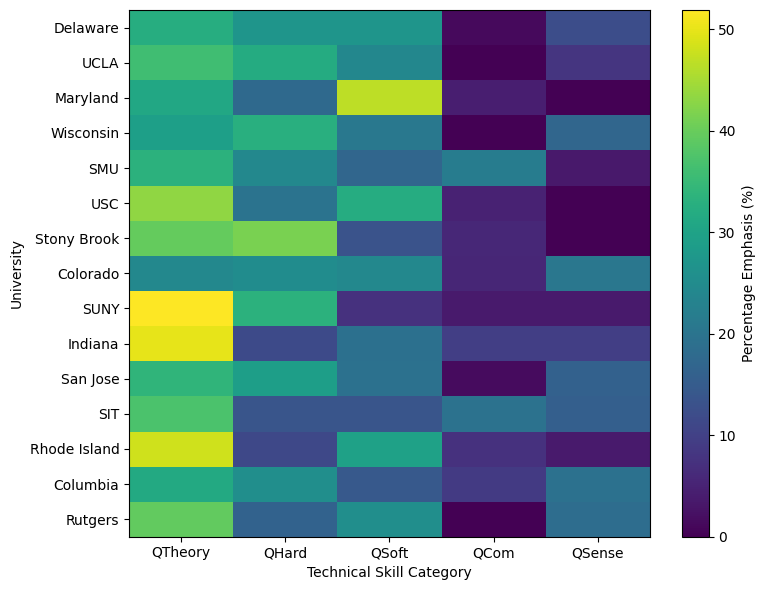}
        \caption*{(a) Skill Emphasis Heatmap}
    \end{minipage}\hfill
    \begin{minipage}{0.55\textwidth}
        
        \includegraphics[width=\linewidth]{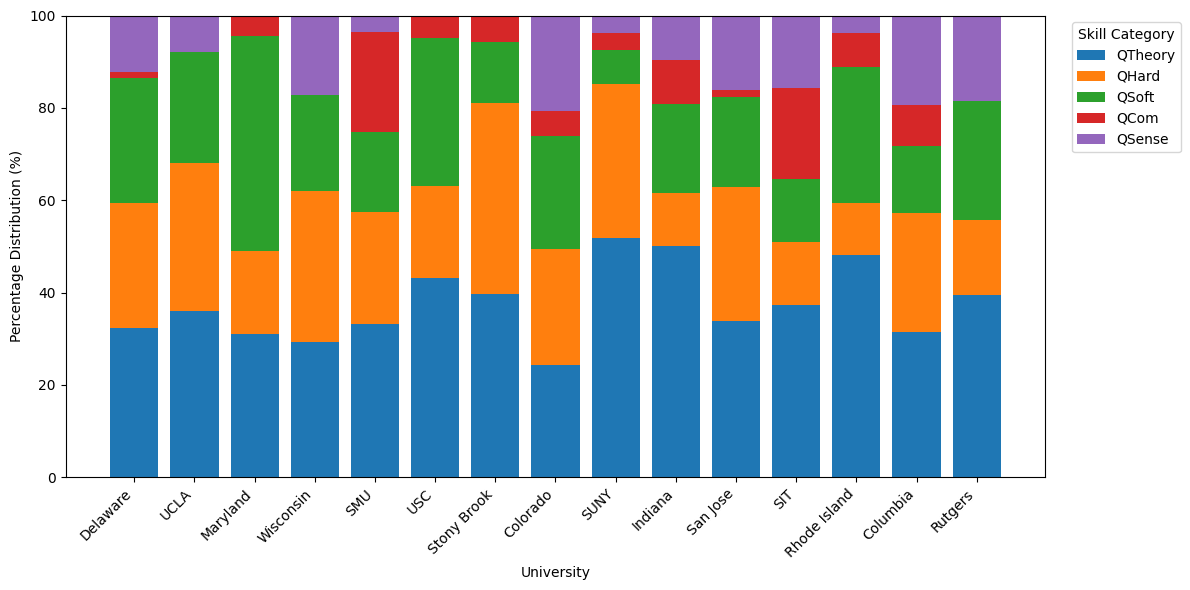}
        \caption*{(b) Stacked Technical Skill Distribution}
    \end{minipage}
    \caption{Cross-program comparison of skill distributions.}
    \label{fig:heatmap-stack}
\end{figure}

Figure 3a presents a heatmap summarizing skill emphasis across all fifteen programs, with rows representing skill categories and columns representing universities. Color intensity reflects the relative proportion of each skill within a program, enabling direct visual comparison of curricular emphasis across institutions. Complementing this view, Figure 3b presents a stacked bar chart of normalized technical skill distributions, with each bar corresponding to a program and segmented by skill category.

Together, these visualizations highlight both commonalities and substantial variation across U.S. primary master’s programs in quantum science and technology. Overall, Figure 3 illustrates a heterogeneous national landscape in which U.S. quantum master’s programs share a common foundation in quantum theory but diverge substantially in how applied, computational, experimental, and specialized skills are integrated. These observed differences provide a descriptive basis for examining alignment with workforce expectations and role frameworks in the subsequent Discussion section.

 \subsection{Applied Learning Opportunities Across Programs}

Figure~4 presents a pie chart summarizing the distribution of formal applied learning opportunities across the fifteen U.S.-based primary master’s programs analyzed. Applied learning was coded based on the presence of explicitly stated experiential components in program descriptions, including industry internships, research-based laboratory experiences, or capstone-style projects.

As shown in Figure~4, one-third of the programs (33.3\%) explicitly incorporate industry internships as part of their curriculum or degree requirements. An additional 33.3\% emphasize research-based laboratory experiences, typically conducted within university research groups or affiliated laboratories. Capstone-style project experiences appear far less frequently, accounting for only 6.7\% of programs. Notably, 26.7\% of programs do not clearly articulate any formal applied learning component in publicly available program materials.

\begin{figure}[H]
  \centering
  \includegraphics[width=12cm]{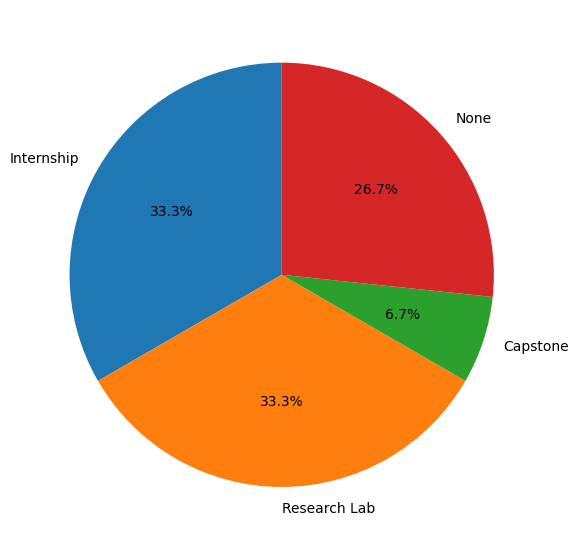}
  \caption{Applied Learning Opportunities}
  \label{fig:applied-learning}
\end{figure}

These results indicate that, while applied learning is present in a majority of programs, its form and emphasis vary substantially. The distribution suggests that experiential preparation is most commonly framed through either industry internships or academic research experiences, with comparatively limited adoption of structured capstone models. The absence of clearly defined applied learning opportunities in over one-quarter of programs highlights variability in how workforce-relevant experiences are embedded within quantum master’s curricula.

\subsection{Discretization of Technical Skills Emphasis and Robustness Analysis}

Figure~5 presents the results of the discretization-based analysis of technical skills emphasis across U.S. primary quantum master’s programs, as introduced in the methodology section. This analysis operationalizes curricular emphasis on technical skill categories using normalized skill frequencies and predefined discretization thresholds. Importantly, the figures represent an \emph{aggregate, system-level distribution} of technical skills across all programs analyzed. Normalized skill counts were pooled across programs prior to discretization to capture the collective training signal conveyed by primary quantum master’s education.

Figure~5a shows a bar graph derived from the baseline discretization thresholds, illustrating the relative emphasis placed on each technical skill category under the primary classification scheme. This visualization provides a snapshot of how technical training is distributed across core quantum-related domains at the system level.

To evaluate the stability and sensitivity of these findings, Figure~5b presents a line graph overlaying results from multiple alternative threshold schemes, ranging from conservative to more liberal discretization choices. By comparing trends across threshold variants, this figure assesses the robustness of observed skill-emphasis patterns to methodological assumptions inherent in threshold selection.

\begin{figure}[H]
    \centering
    \begin{minipage}{0.45\textwidth}
        
        \includegraphics[width=\linewidth]{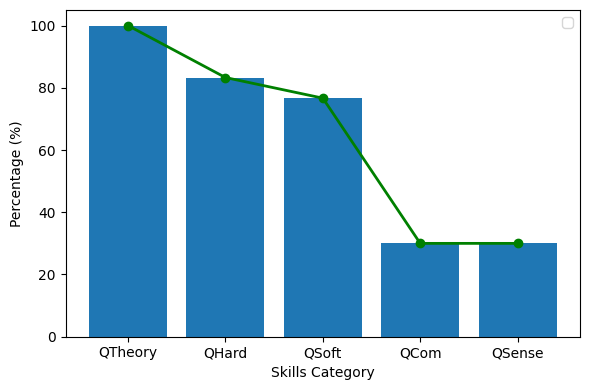}
        \caption*{(a) Technical Skills Distribution}
    \end{minipage}\hfill
    \begin{minipage}{0.45\textwidth}
        
        \includegraphics[width=\linewidth]{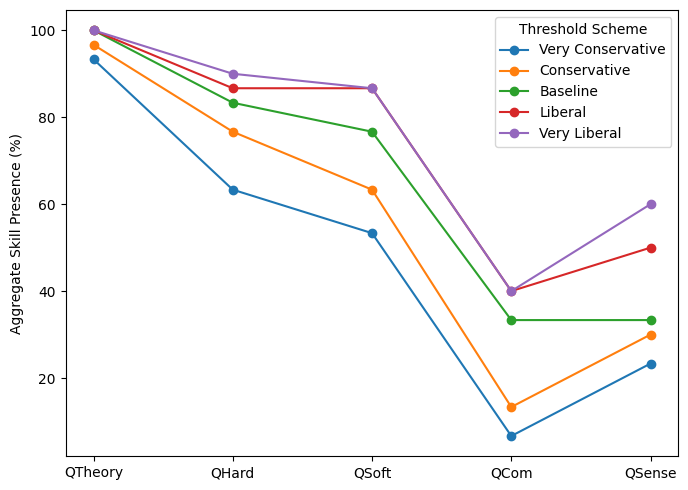}
        \caption*{(b) Sensitivity to Discretization Thresholds}
    \end{minipage}\hfill
    \begin{minipage}{0.55\textwidth}
        
        \includegraphics[width=\linewidth]{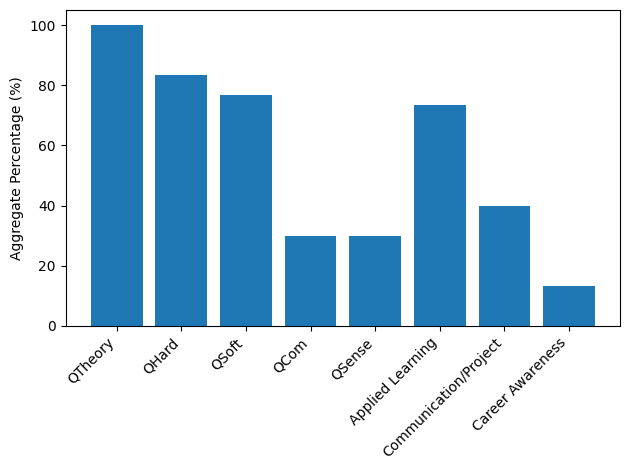}
        \caption*{(c) Technical \& Non-technical Skills Distribution}
    \end{minipage}
    \caption{ Skills Distribution across U.S. Primary MSc Quantum Programs.}
    \label{fig:heatmap-stack}
\end{figure}

Across all threshold schemes, the relative ordering and prominence of major technical skill categories remain largely consistent. This convergence indicates that the aggregate findings are not artifacts of a particular discretization choice, but instead reflect stable patterns in curricular emphasis. The robustness of these results strengthens confidence in the reported distribution of technical skills and supports their interpretation as a meaningful representation of the overall preparation signal provided by U.S. primary quantum master’s programs for the quantum workforce pipeline.

Building on the technical skill aggregation shown in Figure~5a, Figure~5c extends the discretization framework to include both technical and non-technical skill categories. Specifically, Figure~5c presents the aggregate distribution of technical skills alongside applied learning opportunities, communication and project management competencies, and career awareness preparation across U.S. primary quantum master’s programs. This expanded representation provides a holistic system-level profile of curricular emphasis, capturing how programs balance disciplinary training with professional and experiential skill development.

\clearpage
\section{Discussion}

This section interprets the aggregate curricular skill-emphasis patterns identified in this study in relation to established quantum workforce frameworks developed through industry-engaged research. In particular, our analysis is situated within a growing body of quantum workforce studies led by the Rochester Institute of Technology (RIT) and the Colorado-based research consortium, which together provide one of the most comprehensive empirical accounts of non-PhD roles, skill requirements, and workforce pathways in the U.S. quantum ecosystem (Piña et al., 2025, Shams El-Adawy et al., 2025, Shams El-Adawy et al., 2026).

These reports draw on employer interviews, job role analyses, and industry–academic partnerships to characterize the knowledge, skills, and abilities (KSAs) required across a range of quantum roles, with particular attention to quantum-proficient positions accessible to bachelor’s- and master’s-level graduates. Rather than treating workforce needs as static targets, this literature emphasizes role diversity, hybrid skill profiles, and the importance of applied and interdisciplinary preparation alongside technical foundations.

Accordingly, our Discussion adopts an interpretive alignment perspective. Observed curricular emphases are compared with industry-identified role profiles and competency domains to assess the extent to which U.S. primary quantum master’s programs collectively reflect workforce-relevant preparation signals. This approach does not assume a direct mapping between curriculum and employment outcomes, but instead situates curricular patterns within externally articulated workforce expectations. The Discussion proceeds by first examining alignment with industry workforce needs, followed by implications for curriculum design and policy.

\subsection{Alignment with Industry Workforce Needs}

Collectively, the quantum workforce reports produced by the RIT and Colorado research groups articulate four principal role categories that reflect contemporary industry needs: \textit{hardware}, \textit{software}, \textit{bridging}, and \textit{public-facing/business} roles. Hardware roles emphasize the design, fabrication, characterization, and operation of quantum devices; software roles focus on algorithm development, system-level programming, and application deployment; bridging roles span technical and organizational boundaries by translating system requirements across teams; and public-facing or business roles engage in strategy, partnerships, product translation, and stakeholder communication.

To situate our findings within these workforce frameworks without overextending curricular–labor market equivalence, we interpret observed curricular emphases in light of the \textit{quantum-proficient} role level articulated in these reports. This level corresponds closely to the scope of the present study, as it targets roles typically accessible to master’s-level graduates rather than doctoral specialists. Roles at the quantum-proficient level span hardware, software, and externally facing functions, with required knowledge, skills, and abilities (KSAs) that map naturally onto our curriculum coding categories.

Specifically, hardware- and software-oriented quantum-proficient roles align closely with the QHard and QSoft skill domains identified in our analysis. Business-facing and public-facing roles, as well as many bridging roles, place greater emphasis on communication, project coordination, and cross-functional collaboration. In our coding framework, communication and project management competencies are treated as additive to technical preparation rather than as substitutes, reflecting the hybrid skill profiles emphasized in the workforce reports. This distinction aligns with industry characterizations of non-research quantum roles as requiring technical literacy alongside professional and organizational competencies.

Viewed through this lens, our results suggest that U.S. primary quantum master’s programs provide strong foundational preparation for quantum-proficient technical roles, particularly in theory, software, and hardware-adjacent domains. However, variation in the integration of professional and interdisciplinary skill development may influence graduates’ readiness for bridging and public-facing roles emphasized in industry needs assessments. For example, programs with strong emphasis on hardware skills but limited project management or communication training may prepare graduates well for device-focused engineering tasks, while offering less preparation for cross-functional or client-facing responsibilities. Conversely, programs that explicitly integrate professional competencies more closely reflect workforce expectations for roles that span technical, organizational, and strategic domains.

Our findings further align with industry-identified experimental and applied skill needs articulated in the report on \textit{Experimental Skills for Non-PhD Roles in the Quantum Industry}. That study highlights competencies such as instrumentation, experimental control, data analysis, troubleshooting, and systems-level thinking as central to quantum-proficient roles that do not require doctoral training. These competencies correspond closely to the QHard domain and applied components of the QSoft category in our analysis, particularly in programs emphasizing laboratory coursework, hardware-oriented instruction, or project-based implementation. The uneven distribution of these skills across programs observed in our results mirrors the report’s conclusion that access to experimental preparation remains highly program-dependent, reinforcing concerns about variability in workforce readiness for applied and hardware-facing roles.

Qualitative interview findings reported in the Colorado workforce studies further contextualize these curricular patterns. Interviewees consistently noted that, while applicants often demonstrate strong foundational quantum knowledge, weaknesses persist in interdisciplinary preparation, particularly at the intersection of quantum science with business, management, and organizational contexts. This observation aligns with our finding that communication and project management competencies are unevenly integrated across programs and are frequently positioned as supplementary rather than core curricular components. Interviewees also highlighted variability in applicants’ ability to communicate technical concepts to diverse audiences, a concern that directly mirrors our observation that explicit training in communication and professional skills is absent or minimally embedded in a substantial subset of programs.

Finally, industry respondents emphasized the need for stronger coordination between higher education and industry, including expanded applied learning opportunities and clearer communication channels to better align academic preparation with evolving workforce needs. These recommendations resonate with our findings on the heterogeneous availability of internships, capstone projects, and research-based experiential learning across programs. Taken together, the convergence between industry interview insights and observed curricular structures underscores the importance of intentional integration of applied learning, interdisciplinary training, and professional skill development within quantum master’s education.

Importantly, this alignment is interpretive rather than predictive. Our intent is not to infer workforce outcomes directly from curricular structures, but to contextualize observed patterns of skill emphasis within established industry role frameworks.

\subsection{Implications for Curriculum Design and Policy}

The aggregate patterns identified in this study suggest several implications for the design, evaluation, and governance of primary quantum master’s curricula. Rather than pointing to a single optimal curriculum model, the results highlight areas where greater intentionality and coherence could strengthen alignment between educational preparation and the diverse role profiles articulated in contemporary quantum workforce frameworks.

First, the strong and robust emphasis on core technical domains across programs, particularly quantum theory, computation, and hardware-adjacent skills, indicates that U.S. primary quantum master’s programs have largely converged on a shared technical foundation. From a curriculum design perspective, this convergence supports the continued prioritization of rigorous technical instruction as a defining feature of master’s-level quantum education. At the policy level, it suggests that existing program structures are broadly successful in signaling technical competence consistent with quantum-proficient workforce expectations.

At the same time, the observed variability in applied learning opportunities and professional skill integration points to opportunities for curricular enhancement. Figures~4 and~5 show that experiential components such as internships, capstone projects, and structured research experiences are unevenly embedded across programs, and that communication, project management, and career awareness are frequently positioned as supplementary rather than integral elements of the curriculum. For curriculum designers, this pattern suggests that professional and applied competencies may benefit from more explicit curricular framing, either through required coursework, embedded project-based learning, or scaffolded experiential pathways, rather than relying solely on optional or external opportunities.

From a policy perspective, these findings reinforce recommendations from industry-engaged workforce studies calling for stronger coordination between higher education and industry partners. Incentivizing industry–academic collaboration through funding mechanisms, shared training initiatives, or co-designed experiential modules could help reduce variability in access to applied learning while preserving institutional flexibility. Importantly, such coordination need not imply uniform curricula, but rather clearer articulation of how programs prepare students for different segments of the quantum workforce ecosystem.

The results also suggest implications for how quantum master’s programs communicate their objectives and outcomes to students and external stakeholders. The diversity of role profiles emphasized in workforce reports implies that no single program can—or should—optimize preparation for all quantum roles simultaneously. Explicitly signaling curricular strengths, such as emphasis on hardware experimentation, software development, or interdisciplinary translation, may help align student expectations with program outcomes and support more informed pathways into the workforce.

The observed diversity in curricular strengths across quantum master’s programs may not solely reflect misalignment with workforce needs, but may also indicate intentional or emergent specialization toward distinct segments of the quantum sector, such as hardware development, quantum software, quantum information theory, or applications in sensing and communication. From this perspective, variation across programs could be understood as supporting a heterogeneous workforce rather than a uniform training model. Nevertheless, such specialization underscores the importance of clearly articulating and empirically understanding the skill requirements associated with different quantum career pathways. Without explicit connections between program curricula and sector-specific workforce demands, students may not be adequately prepared for the roles these programs implicitly target

An additional implication concerns the role of \emph{career awareness} as a distinct and consequential component of workforce preparation. While technical competence and applied experience are necessary for entry into the quantum workforce, industry reports consistently emphasize that students’ understanding of available role pathways, organizational structures, and hiring practices significantly shapes their ability to translate preparation into employment. Our results indicate that explicit career awareness elements, such as exposure to quantum role taxonomies, industry speakers at departmental seminars, workforce-aligned advising, or structured guidance on non-academic career trajectories, are among the least consistently articulated components across programs (Figure~5c). For students entering an emerging and heterogeneous workforce, limited visibility into role diversity, skill expectations, and career progression may constrain informed decision-making and weaken the signaling value of otherwise strong technical preparation.

From a curriculum and policy perspective, this finding suggests that career awareness should not be treated solely as an extracurricular or advising function, but as a curricular design consideration closely linked to workforce alignment. Embedding career-oriented content within coursework, seminars, or capstone experiences, particularly content that reflects industry-defined role profiles and hybrid skill expectations, may help students more effectively contextualize their training within real-world employment pathways. Such integration aligns with workforce study recommendations emphasizing transparency of role expectations and early exposure to industry norms, and may be especially impactful for master’s-level students seeking timely entry into quantum-proficient roles.

Taken together, these implications suggest that future efforts in quantum curriculum design and policy should focus less on redefining technical foundations, which appear broadly stable, and more on strategically integrating applied, interdisciplinary, career awareness, and professional competencies in ways that reflect the heterogeneous and evolving nature of the quantum workforce.

Finally, at a systems level, the aggregate perspective adopted in this study highlights the value of viewing quantum workforce preparation as a distributed educational ecosystem rather than a set of isolated programs.

\clearpage
\section{Conclusion and Future Direction}

This study provides a systematic, program-level analysis of how U.S. primary quantum master’s programs distribute curricular emphasis across core quantum workforce skill categories. By aggregating course offerings and coding them against industry-informed skill domains, we identify both areas of strong alignment, particularly in theoretical foundations and emerging software, hardware competencies, and persistent gaps in applied training, career awareness, and non-technical professional preparation. These findings suggest that while many programs are succeeding in building conceptual and technical depth, alignment with the heterogeneous demands of the quantum workforce remains uneven.

Importantly, the aggregate nature of this analysis highlights structural trends across programs rather than the learning experiences of students. As such, several future research directions emerge. First, a closer qualitative examination of applied learning opportunities is warranted. This includes investigating the nature and depth of hands-on training students receive, such as laboratory access, hardware interaction, software toolchains, industry-embedded projects, and experiential learning components. Understanding how applied skills are taught, not simply whether such courses exist, would provide a more nuanced picture of workforce readiness.

Second, similar curriculum alignment analyses should be extended to quantum certificate programs, micro-credentials, and quantum specializations housed within broader physics, engineering, or computer science degrees. As these pathways are increasingly marketed as rapid entry points into the quantum workforce, it is critical to assess whether they meaningfully prepare graduates for transitions into quantum roles, particularly for learners without prior quantum backgrounds.

Finally, scaling this work will require incorporating richer program-level and student-level data. Future studies could integrate detailed course syllabi, instructional practices, student interviews, and graduate employment outcomes to triangulate curricular intent with educational practice and workforce placement. Such mixed-methods approaches would strengthen the evidence base needed to inform curriculum design, institutional policy, and national quantum workforce development strategies.

Together, these directions point toward a more comprehensive, evidence-driven understanding of how quantum education pathways can be designed to support timely, equitable, and effective entry into the evolving quantum workforce.
\section{Acknowledgements}

This work was conducted as part of the IBM Quantum Qiskit Advocate Mentorship Program.

\clearpage
\section*{}


\end{document}

%% file: Text/0.1-Big_Figures.tex
\begin{figure}[p]
\centering

\begin{subfigure}{0.3\textwidth}
\centering
\includegraphics[width=\linewidth,height=0.22\textheight,keepaspectratio]{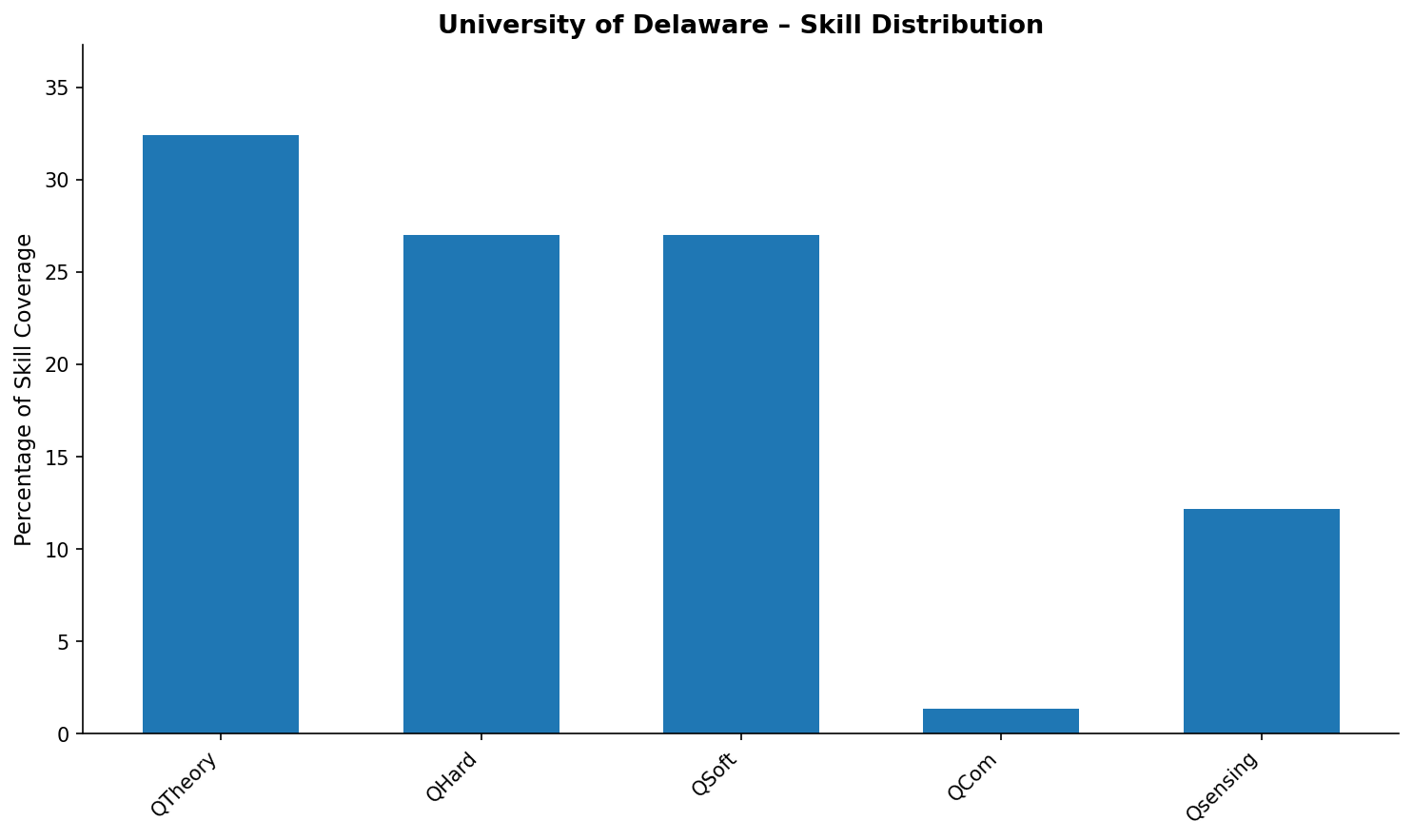}
\caption{Delaware}
\end{subfigure}
\hfill
\begin{subfigure}{0.3\textwidth}
\centering
\includegraphics[width=\linewidth,height=0.22\textheight,keepaspectratio]{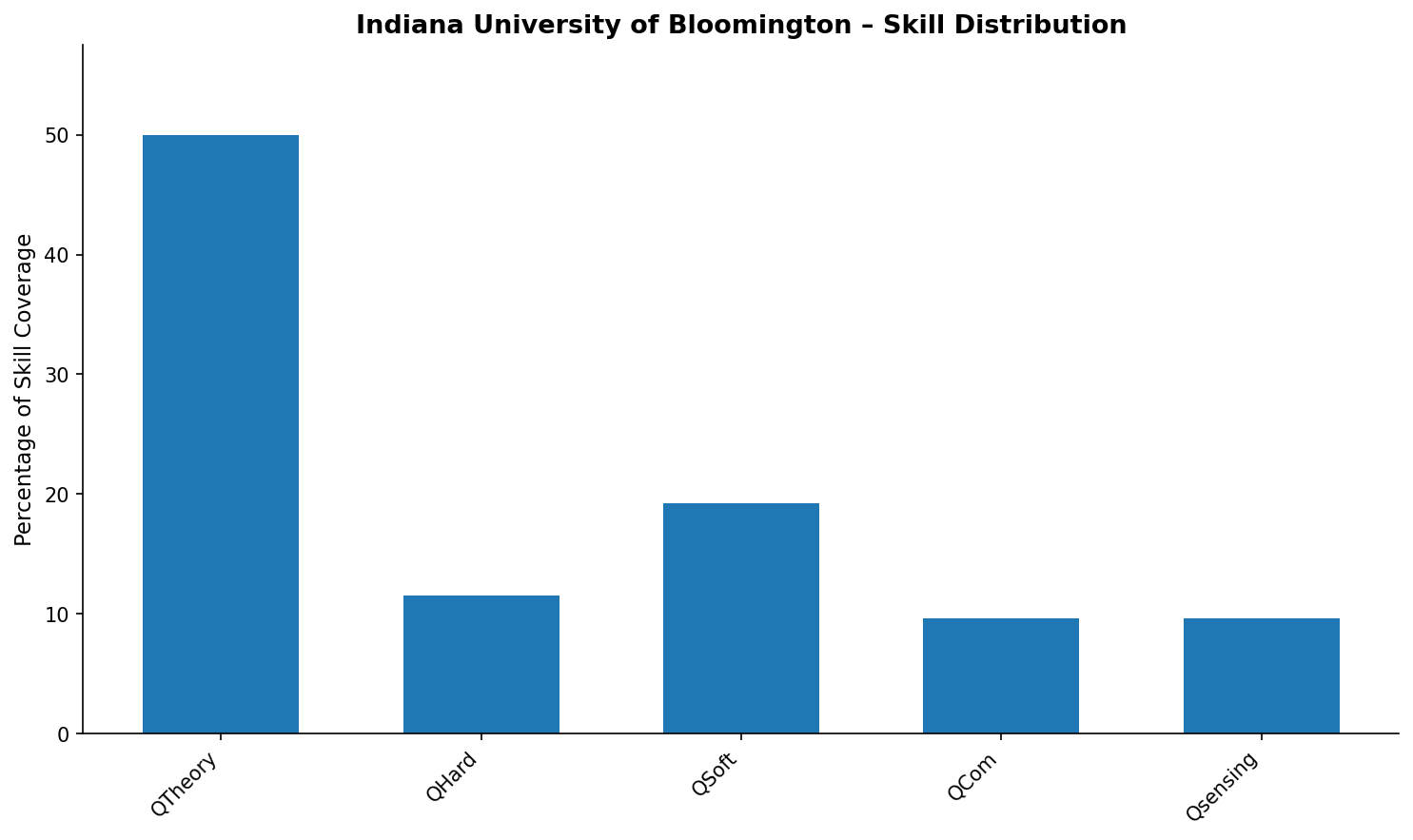}
\caption{Indiana}
\end{subfigure}
\hfill
\begin{subfigure}{0.3\textwidth}
\centering
\includegraphics[width=\linewidth,height=0.22\textheight,keepaspectratio]{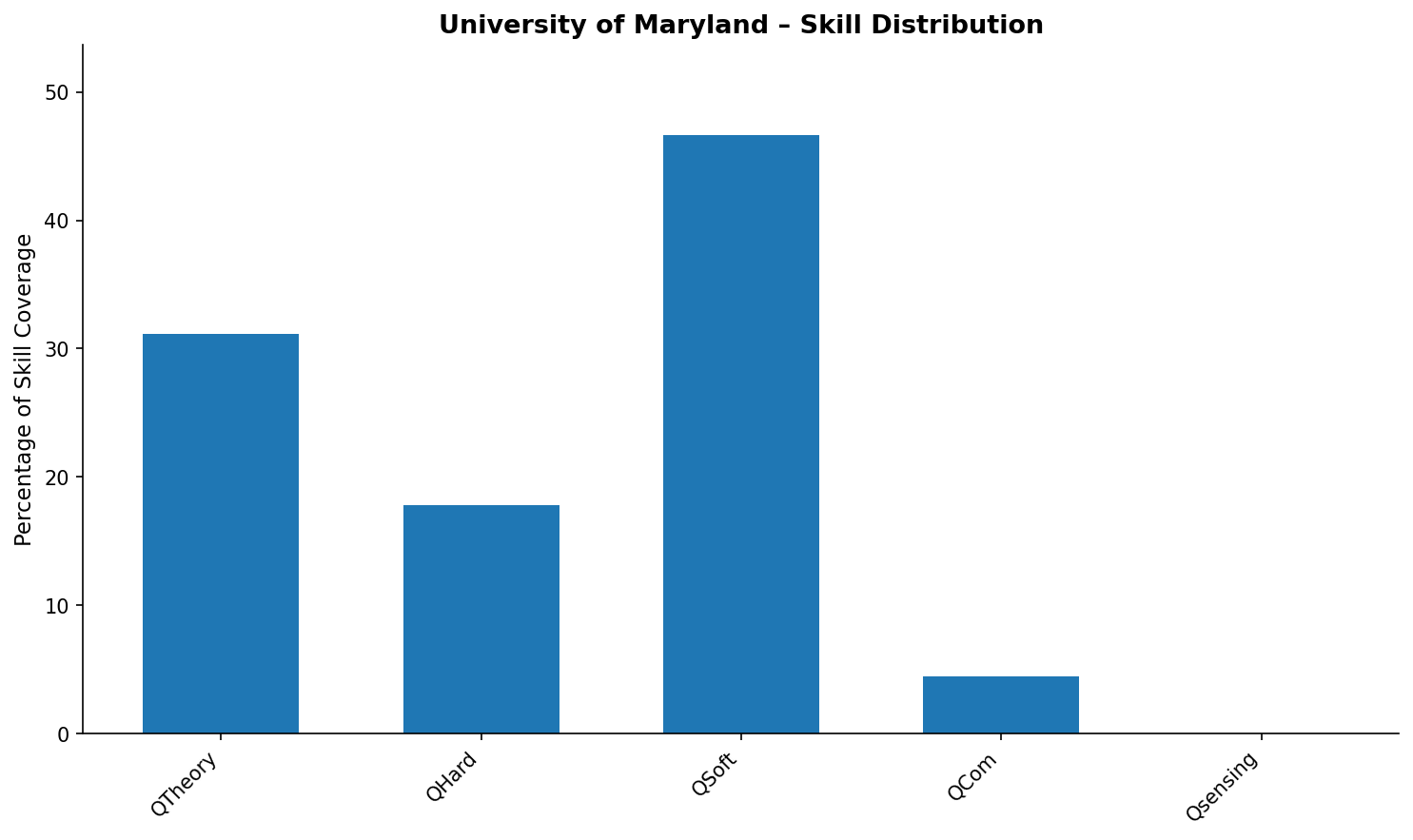}
\caption{Maryland}
\end{subfigure}

\vspace{0.4em}

\begin{subfigure}{0.3\textwidth}
\centering
\includegraphics[width=\linewidth,height=0.22\textheight,keepaspectratio]{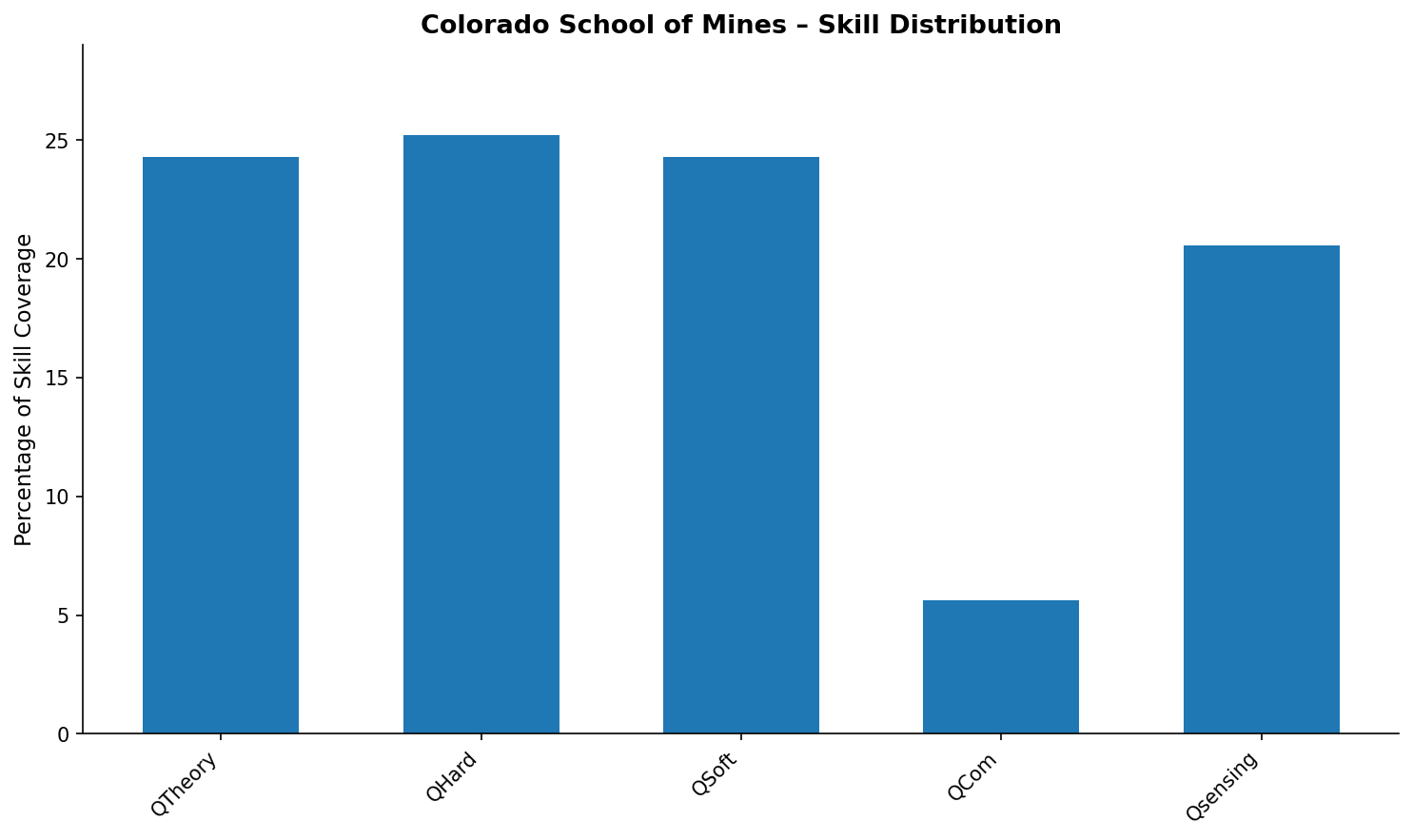}
\caption{Colorado}
\end{subfigure}
\hfill
\begin{subfigure}{0.3\textwidth}
\centering
\includegraphics[width=\linewidth,height=0.22\textheight,keepaspectratio]{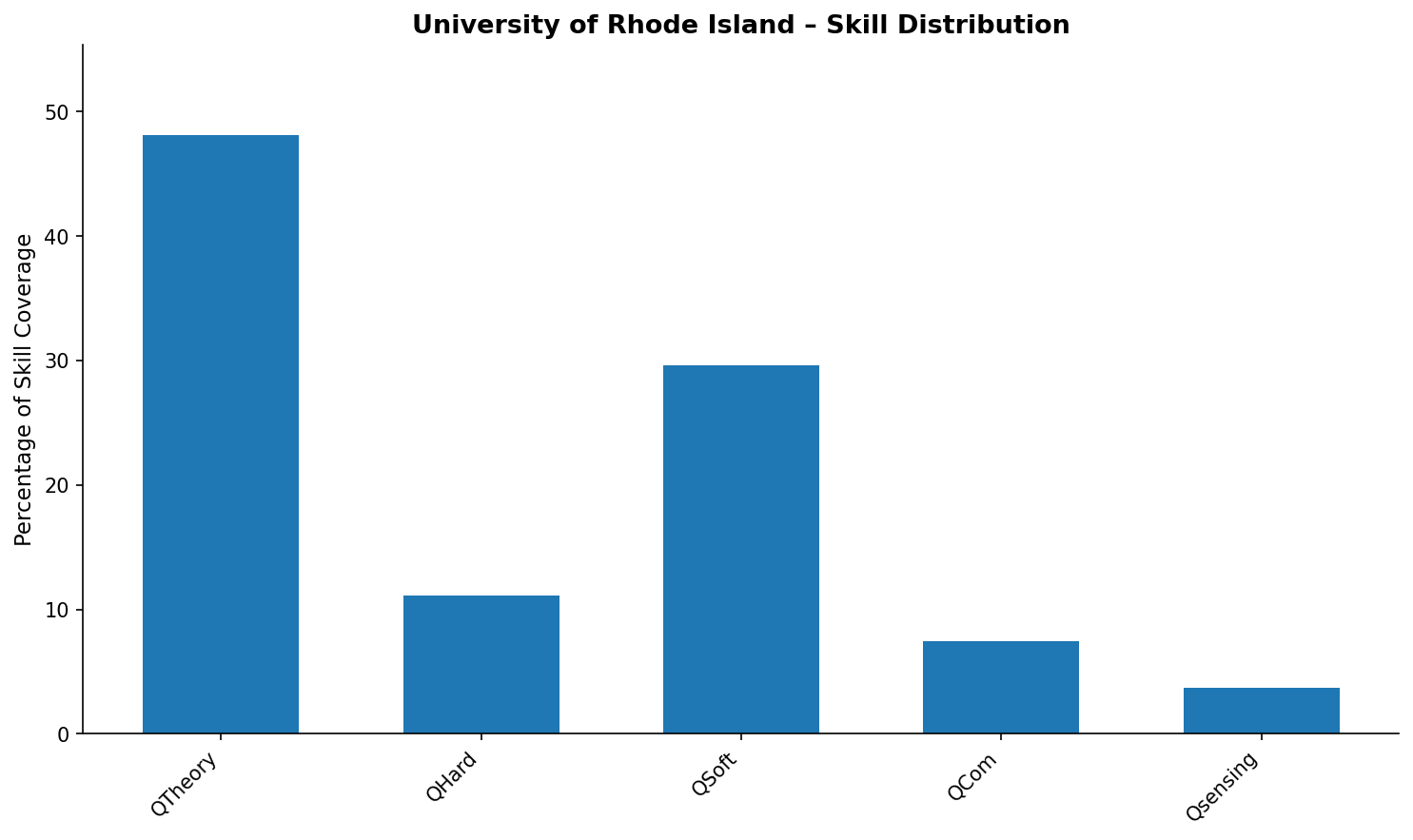}
\caption{Rhode Island}
\end{subfigure}
\hfill
\begin{subfigure}{0.3\textwidth}
\centering
\includegraphics[width=\linewidth,height=0.22\textheight,keepaspectratio]{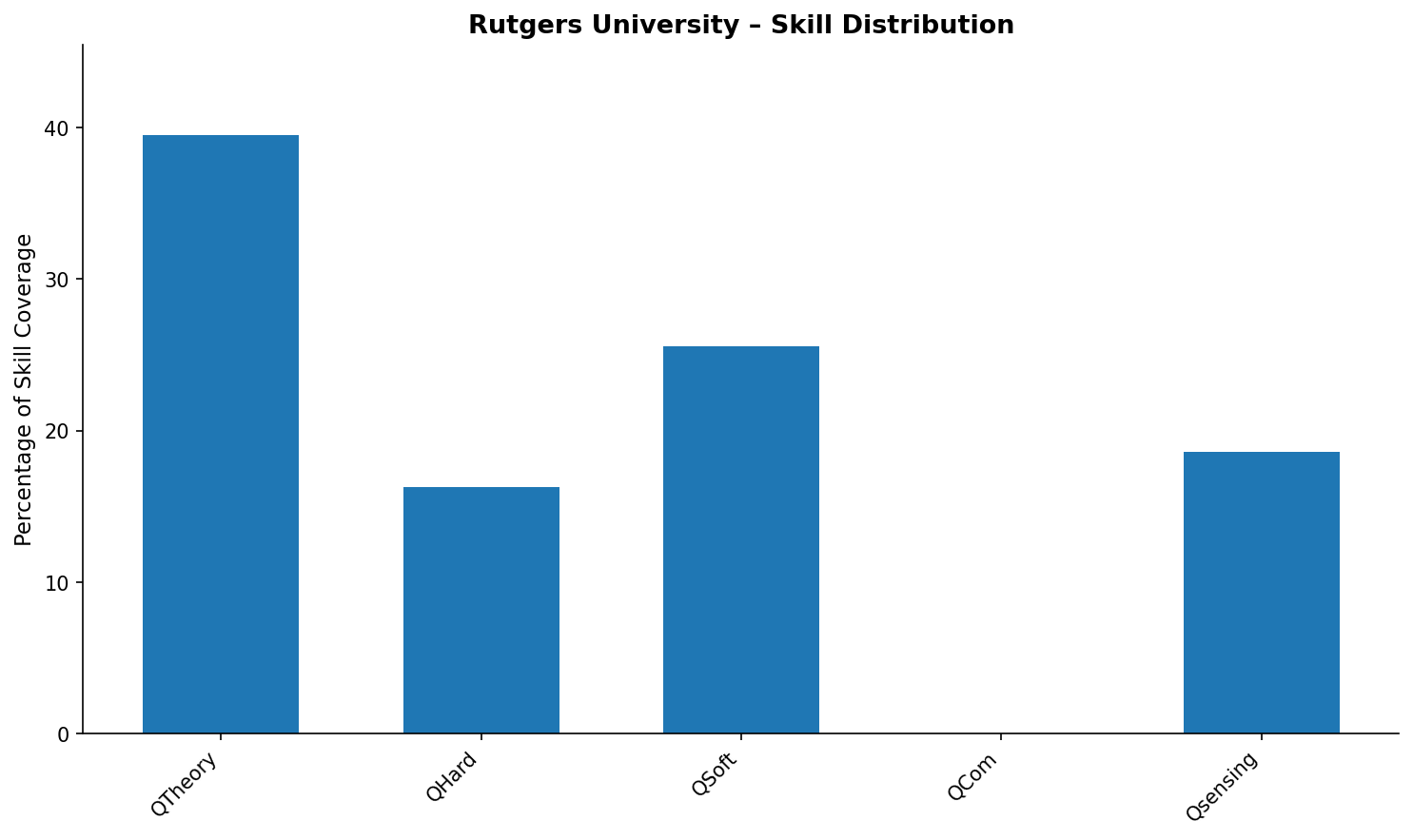}
\caption{Rutgers}
\end{subfigure}

\vspace{0.4em}

\begin{subfigure}{0.3\textwidth}
\centering
\includegraphics[width=\linewidth,height=0.22\textheight,keepaspectratio]{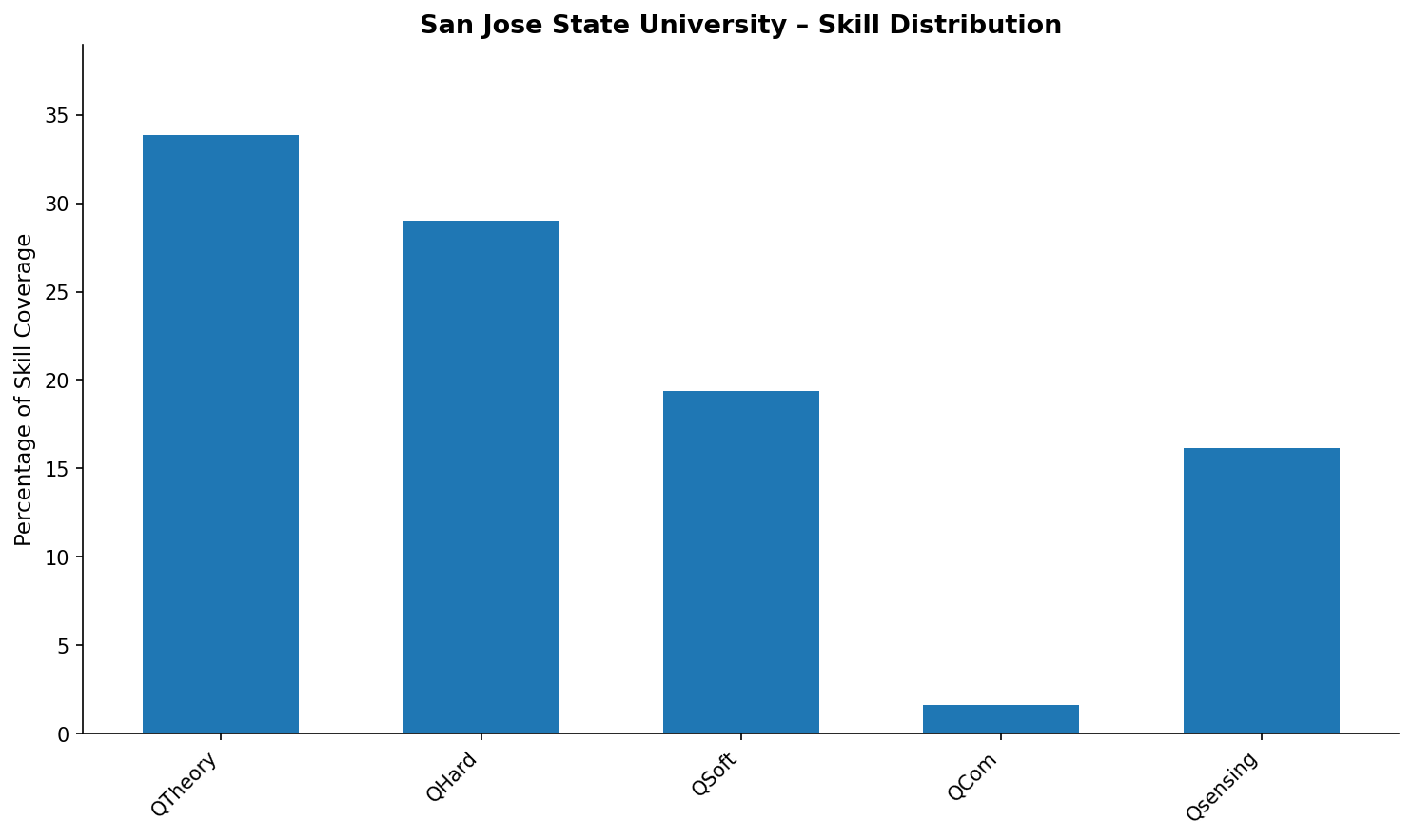}
\caption{San Jose}
\end{subfigure}
\hfill
\begin{subfigure}{0.3\textwidth}
\centering
\includegraphics[width=\linewidth,height=0.22\textheight,keepaspectratio]{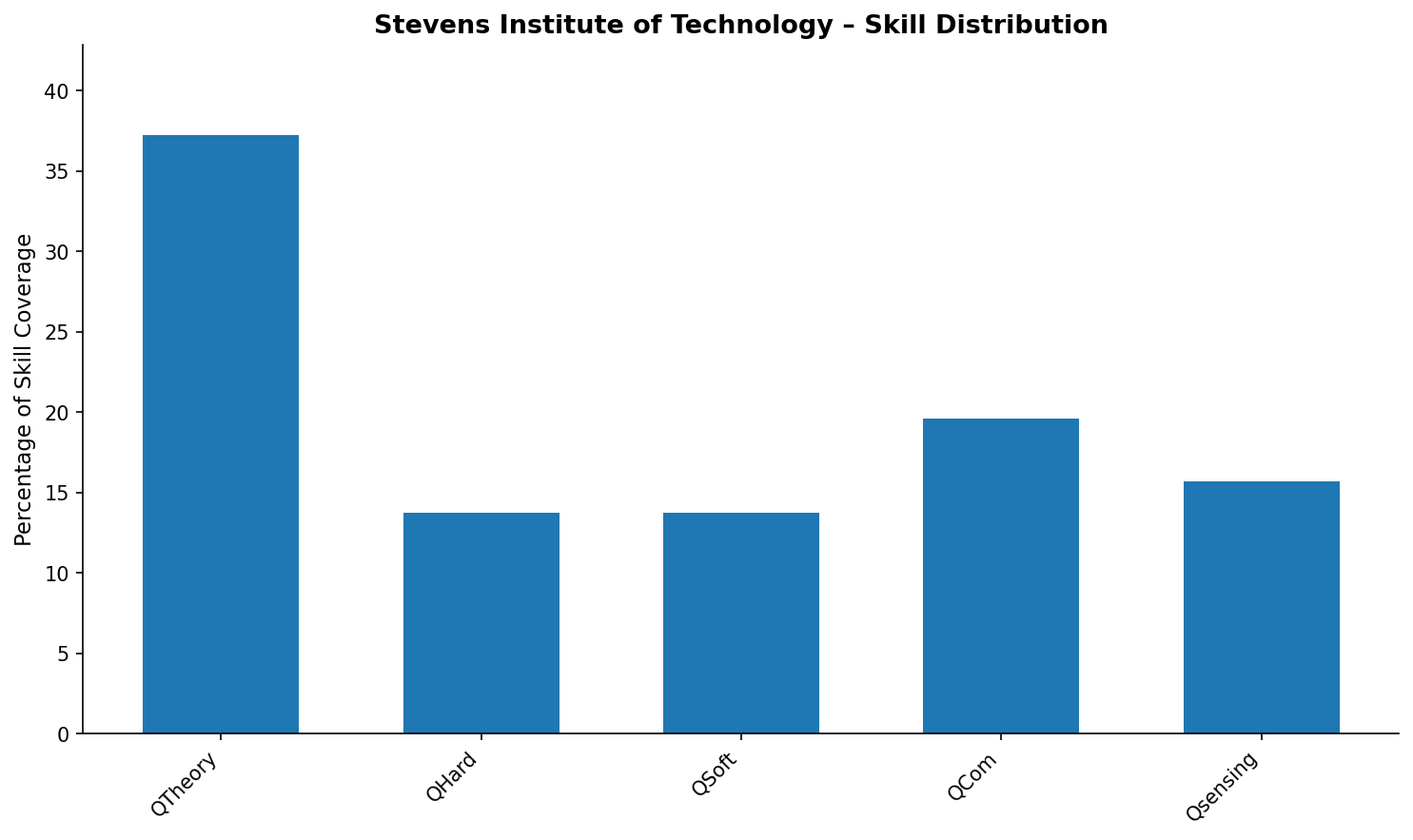}
\caption{SIT}
\end{subfigure}
\hfill
\begin{subfigure}{0.3\textwidth}
\centering
\includegraphics[width=\linewidth,height=0.22\textheight,keepaspectratio]{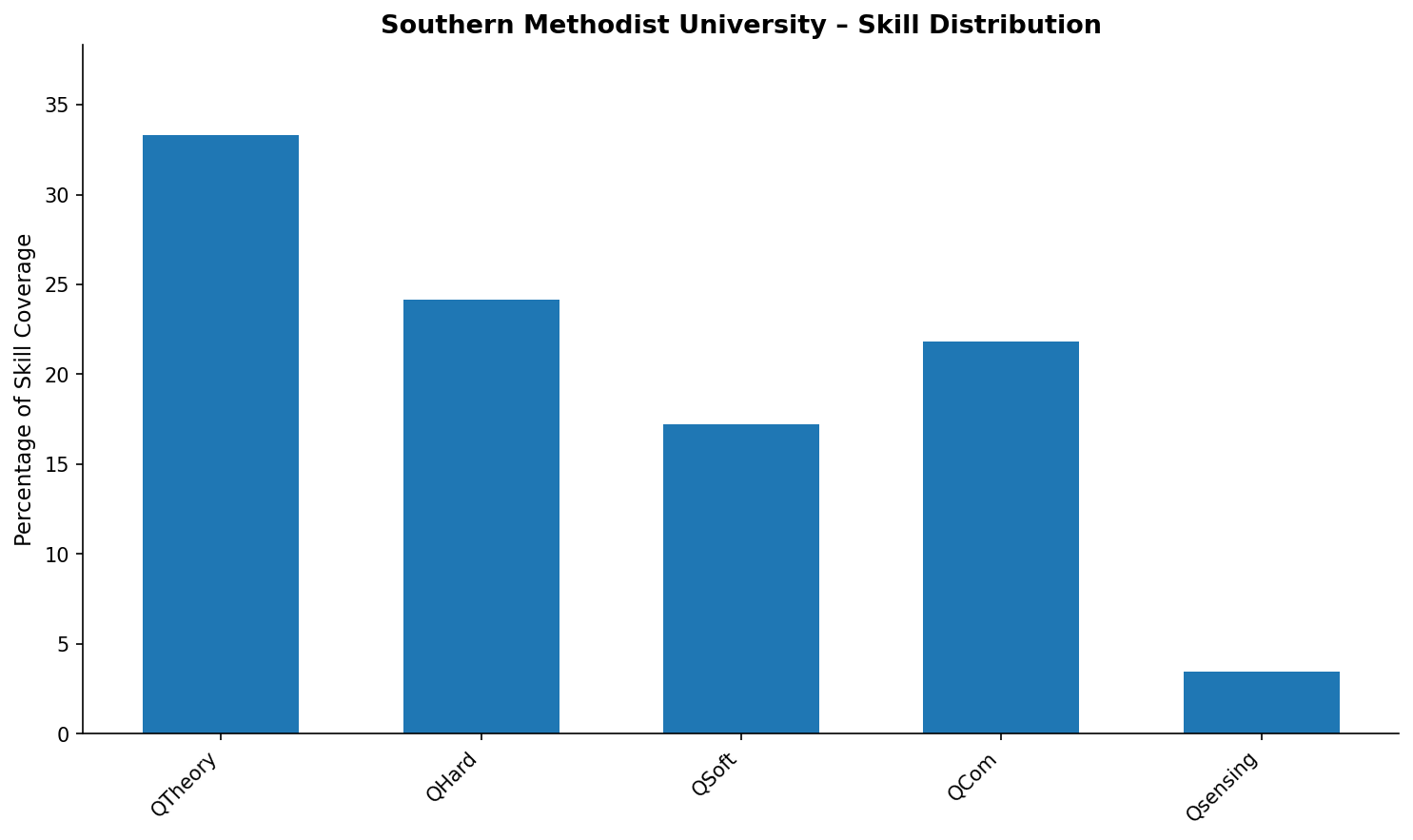}
\caption{SMU}
\end{subfigure}

\vspace{0.4em}

\begin{subfigure}{0.3\textwidth}
\centering
\includegraphics[width=\linewidth,height=0.22\textheight,keepaspectratio]{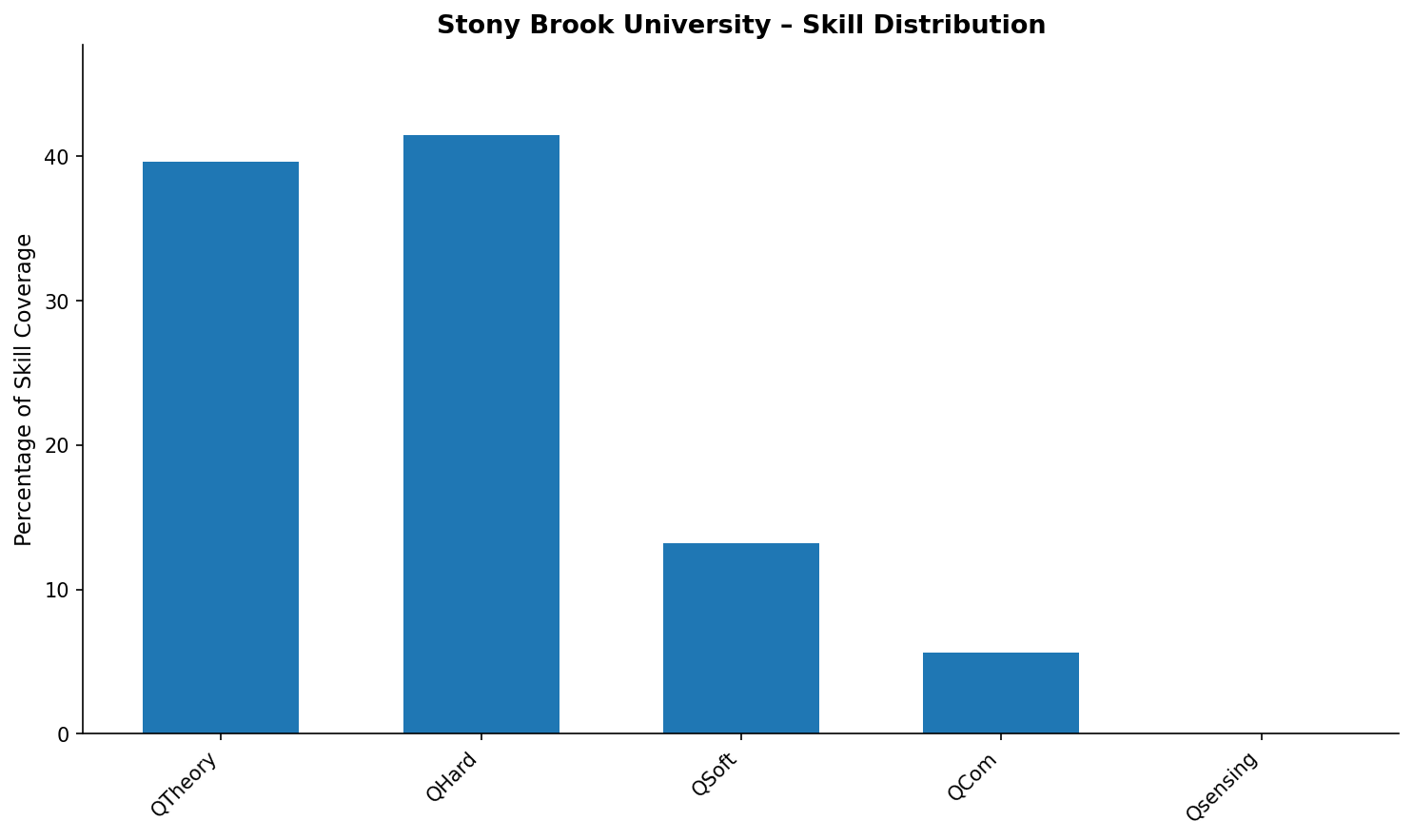}
\caption{Stony Brook}
\end{subfigure}
\hfill
\begin{subfigure}{0.3\textwidth}
\centering
\includegraphics[width=\linewidth,height=0.22\textheight,keepaspectratio]{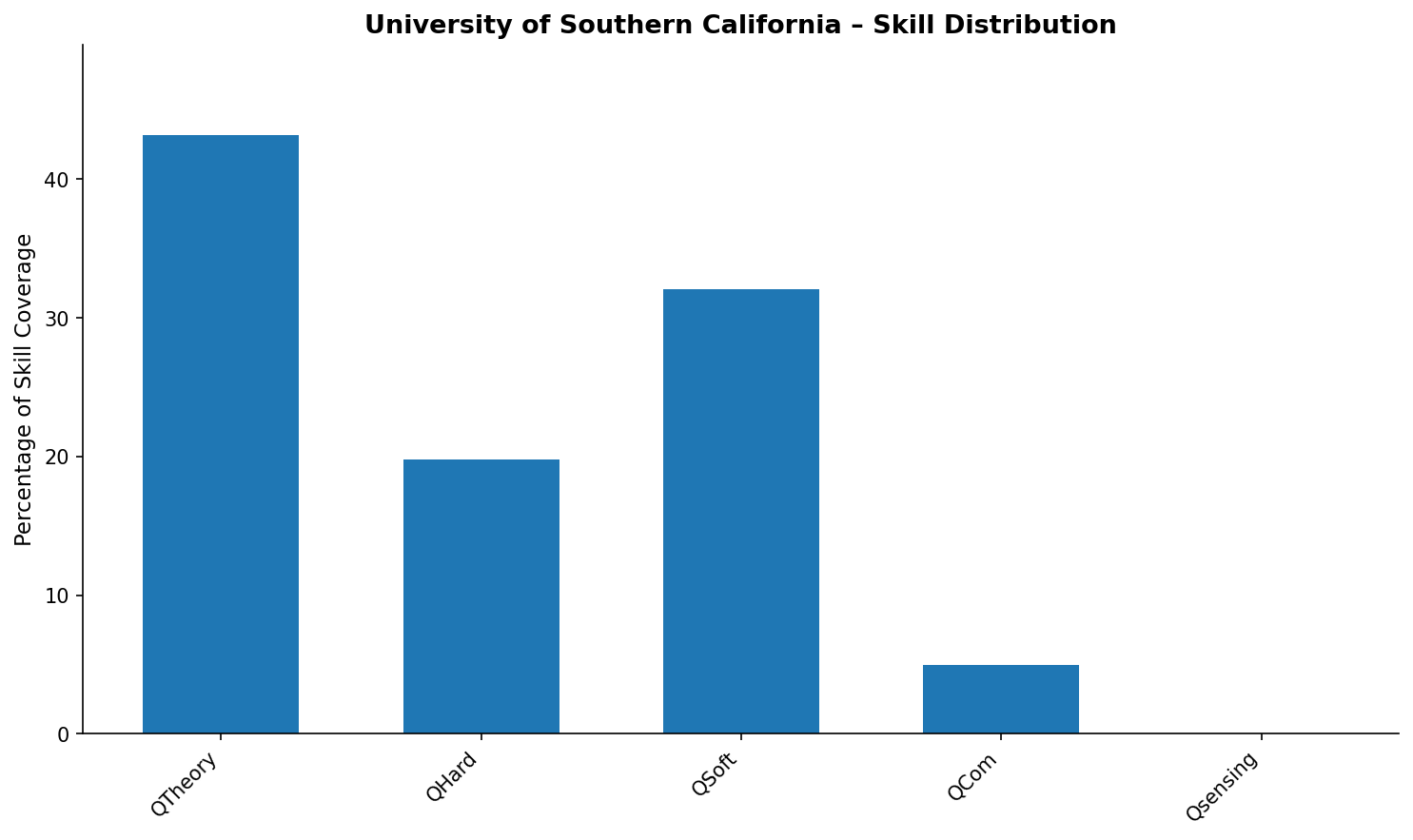}
\caption{USC}
\end{subfigure}
\hfill
\begin{subfigure}{0.3\textwidth}
\centering
\includegraphics[width=\linewidth,height=0.22\textheight,keepaspectratio]{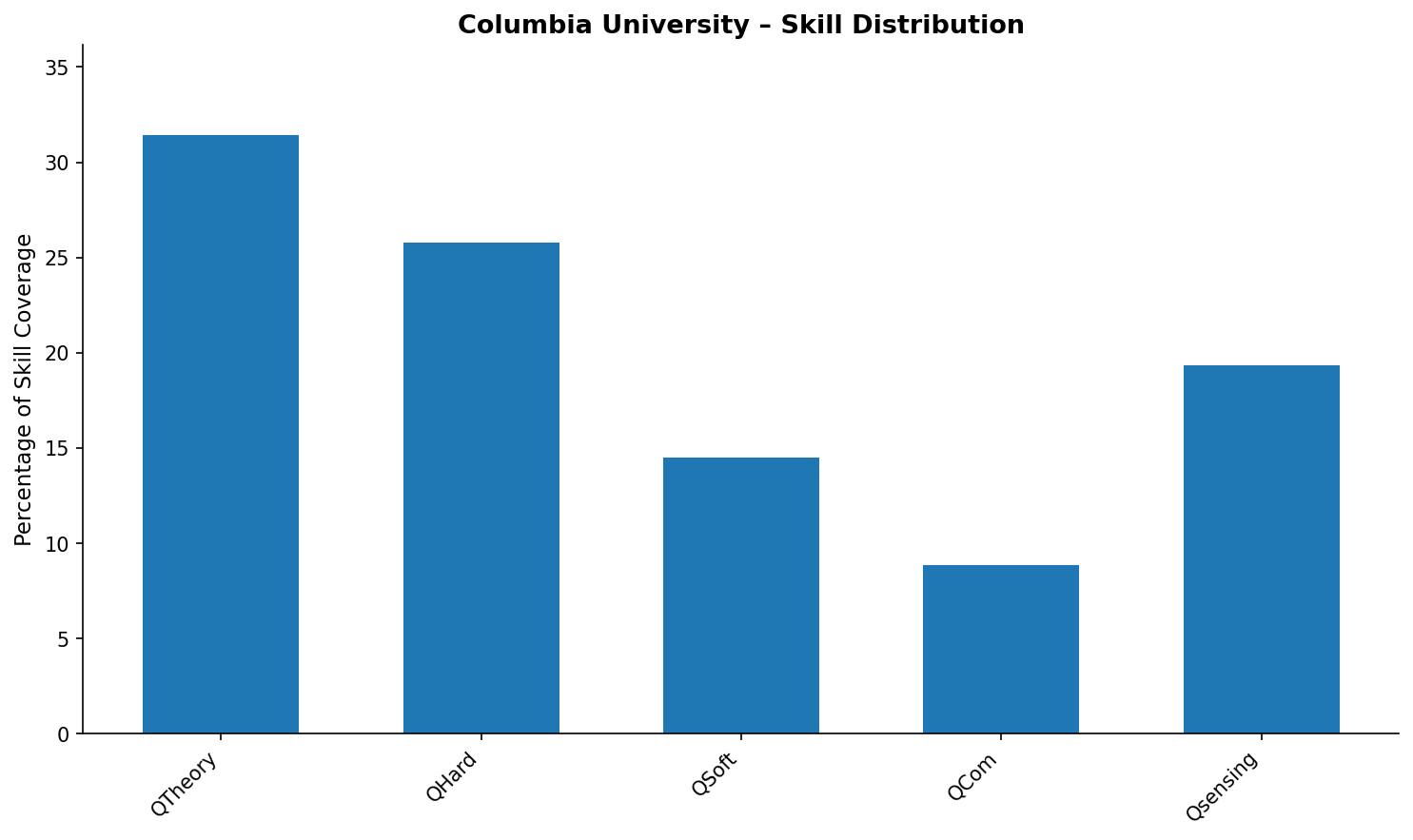}
\caption{Columbia}
\end{subfigure}

\vspace{0.4em}

\begin{subfigure}{0.3\textwidth}
\centering
\includegraphics[width=\linewidth,height=0.22\textheight,keepaspectratio]{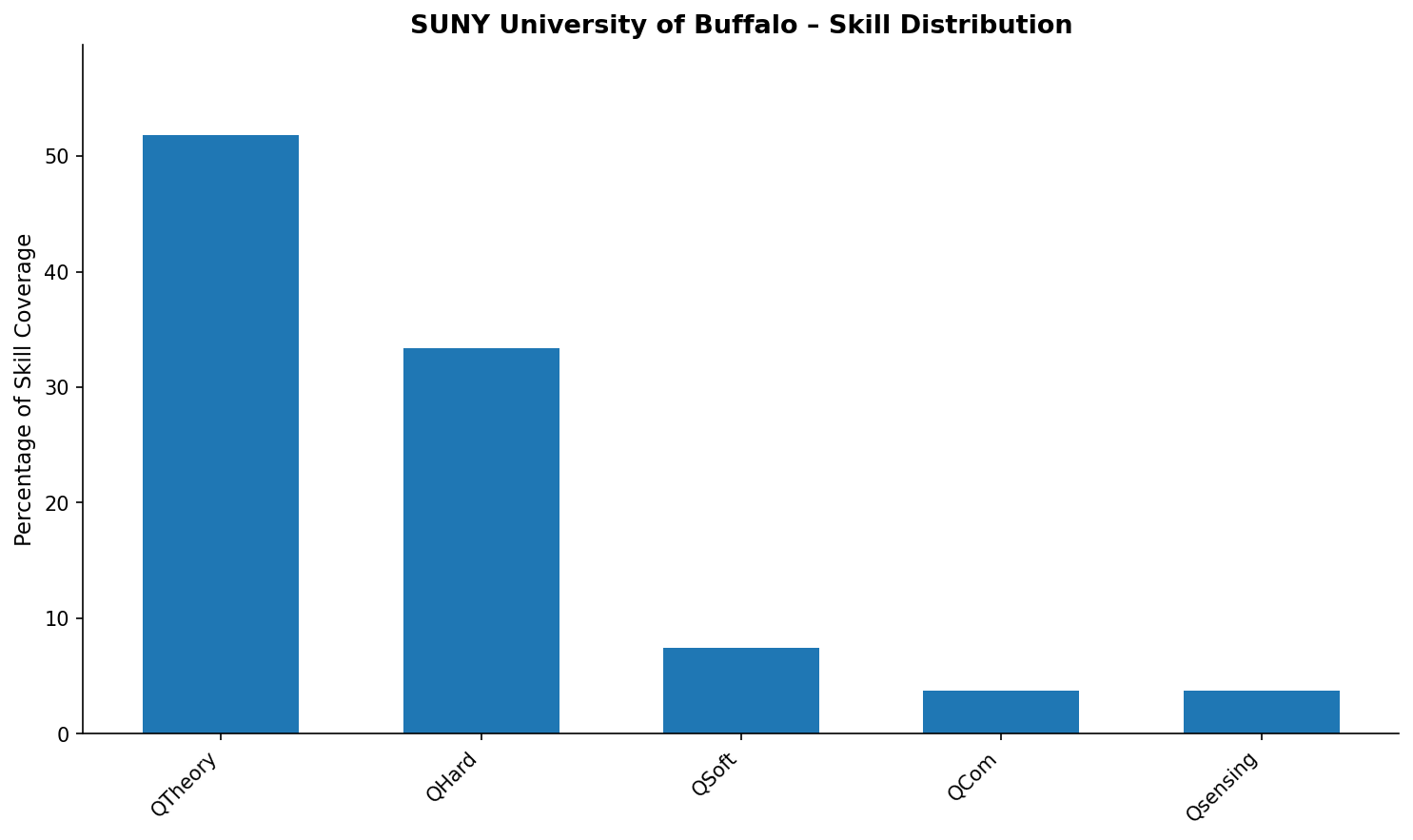}
\caption{SUNY}
\end{subfigure}
\hfill
\begin{subfigure}{0.3\textwidth}
\centering
\includegraphics[width=\linewidth,height=0.22\textheight,keepaspectratio]{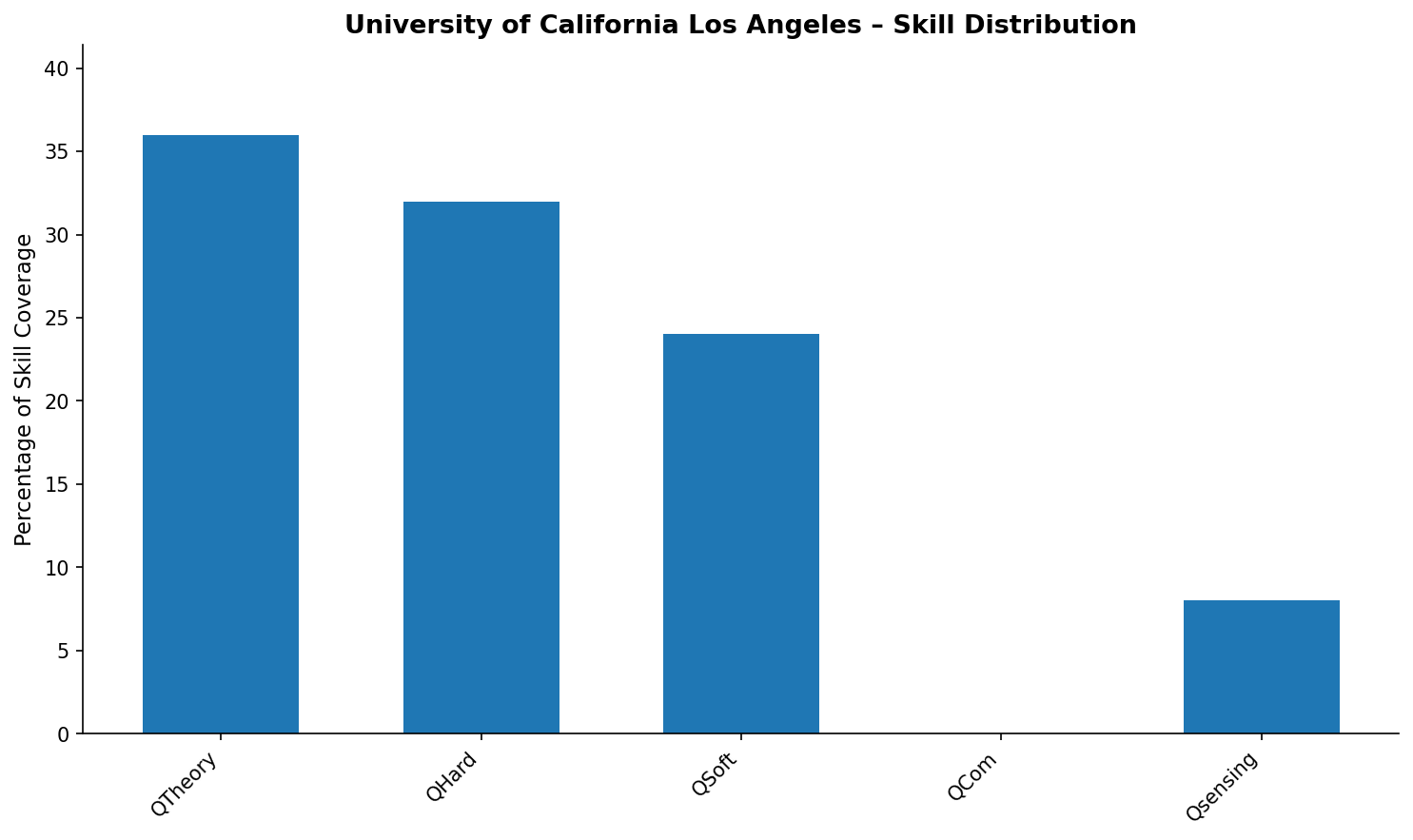}
\caption{UCLA}
\end{subfigure}
\hfill
\begin{subfigure}{0.3\textwidth}
\centering
\includegraphics[width=\linewidth,height=0.22\textheight,keepaspectratio]{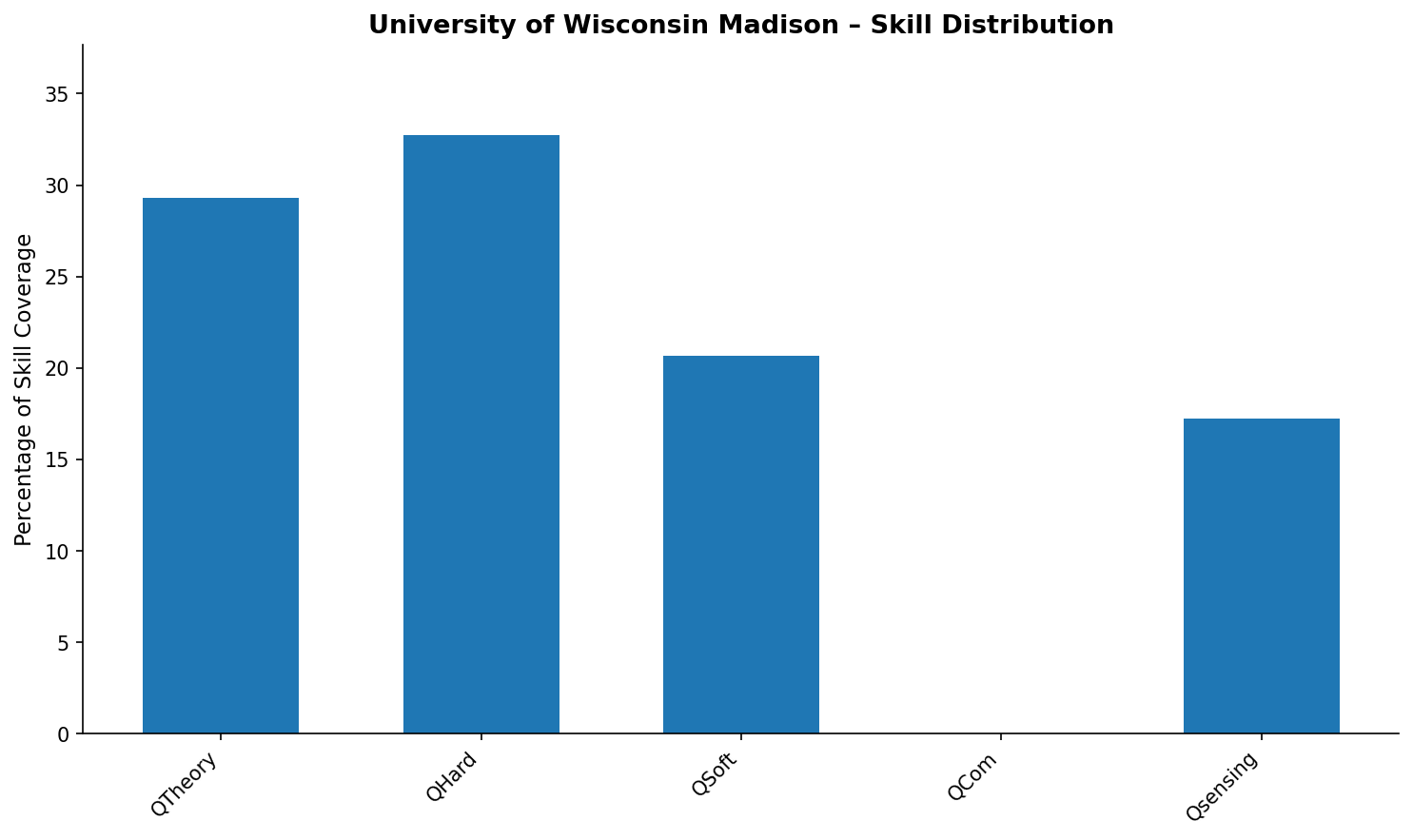}
\caption{Wisconsin}
\end{subfigure}

\caption{Skill distribution per quantum master’s program.}
\label{fig:skill-distribution}
\end{figure}